\documentclass[runningheads]{llncs}

\usepackage[T1]{fontenc}
\usepackage{amsmath,amssymb}
\usepackage{booktabs}
\usepackage{graphicx}
\usepackage{xurl}
\usepackage[hidelinks]{hyperref}
\usepackage{alloy}
\usepackage{beramono}

\newcommand{\methodname}[1]{\textup{\textsf{#1}}}
\newcommand{\arbitrarysplit}{\methodname{Arbitrary Split}}
\newcommand{\balancedsplit}{\methodname{Balanced Split}}
\newcommand{\jointcover}{\methodname{Joint Cover}}
\newcommand{\enumeratedcover}{\methodname{Enumerated Cover}}
\newcommand{\methodheading}[2]{\shortstack{\methodname{#1}\\\methodname{#2}}}

\begin{document}

\title{Symbolic Synthesis of Distinguishing Test Batches for Specification Narrowing}
\titlerunning{Synthesis of Distinguishing Test Batches}

\author{Alcino Cunha \orcidID{0000-0002-2714-8027} \and
Nuno Macedo\orcidID{0000-0002-4817-948X}}
\authorrunning{A. Cunha and N. Macedo}
\institute{INESC TEC \& Universidade do Minho \\ 
Braga, Portugal \\ 
\email{\{alcino,nmacedo\}@di.uminho.pt}}

\maketitle

\begin{abstract}
Ambiguous requirements can admit several non-equivalent formal specifications.
We study the symbolic synthesis of complete distinguishing test batches for
specification narrowing through expert review. Synthesized upfront, these fixed
suites of instances or traces separate every candidate pair and support
asynchronous and parallel classification. We formalize
batch synthesis as a candidate-pair cover problem over an implicit test space
and develop four methods for bounded Alloy models and temporal traces. Two
cover-based methods obtain minimum-cardinality batches through joint MaxSAT
optimization or SAT enumeration followed by explicit minimum-cover selection.
Two tree-guided methods use SAT or balanced MaxSAT and reuse tests across
branches. Our evaluation covers 196 static and 32 temporal configurations.
On paired static configurations, reuse reduces balanced batch cardinality by
42\% relative to the unreused construction tree, and minimum covers are a
further 29\% smaller. The SAT-based tree-guided method completes every
main-study configuration within 0.43 seconds, while the exact methods exhibit
different scalability profiles, particularly for temporal models. Complete
batches also support adaptive review, with balanced batches nearly attaining the
theoretical lower bound on mean classification cost. On completed benchmarks,
modest additional upfront synthesis effort provides complete reusable suites
that also support competitive adaptive classification.

\end{abstract}

\section{Introduction}
\label{sec:introduction}

Developing a formal model may require choosing among alternative
specifications because weak or ambiguous requirements admit multiple
interpretations. Candidates may be written by different experts or produced
automatically. Program synthesis and repair, for example, may return several
artifacts consistent with incomplete input/output
examples~\cite{shriver2017end}; related techniques synthesize temporal
specifications from examples~\cite{gavran2020interactive,zhang2025constrained}
and repair relational
specifications~\cite{wang2019arepair,brida2021beafix,gutierrez2022icebar,cerqueira2022timely}.
More recently, LLMs have been used to generate formal specifications from
natural language, a process known as
\emph{autoformalization}~\cite{weng2025autoformalization,mendoza2026automating}.
Consistency with available examples does not establish which candidate
captures the intended requirement, and generating specifications from
potentially ambiguous natural language does not itself provide an independent
semantic oracle. Choosing among the candidates may therefore require
additional evidence from a domain expert, who may find concrete scenarios
easier to judge than the formal specifications themselves.

Specification narrowing obtains this evidence through concrete tests:
instances or execution traces on which candidates disagree. An expert
classifies a test as desirable or undesirable, eliminating candidates whose
predictions disagree with that judgment. In this paper, we study the synthesis
of a \emph{complete distinguishing test batch} for a given candidate set. A
batch is complete when every candidate pair disagrees on at least one of its
tests. Equivalently, each candidate has a unique pattern of acceptance and
rejection across the batch. Classifying the complete batch therefore leaves
at most one candidate consistent with the expert's judgments.

All tests in a batch are synthesized in advance of expert review. This
supports workflows in which tests are distributed to reviewers, classified
asynchronously or in parallel, or retained as a common basis for comparing
and revisiting judgments. The batch also makes the complete review workload
explicit upfront: its cardinality is the number of classifications required
for full review and an upper bound for adaptive narrowing. Existing online
narrowing approaches~\cite{shriver2017end,gavran2020interactive,mendoza2026automating,prasad2026pick}
synthesize tests in an interactive session, using each classification to guide
the next synthesis query. Complete batch synthesis must instead provide sufficient distinctions
for every possible intended candidate before any answers are available.

The central challenge is to synthesize a small complete batch at reasonable
computational cost. Each feasible test separates some candidate pairs, and a
complete batch must collectively cover all pairs. Given an explicit pool of
tests, minimizing batch cardinality is an instance of the classical NP-hard
Minimum Test Set problem~\cite{GareyJohnson,moret1985minimizing}. In our context, however,
the test space is represented implicitly by a logical theory and analysis
bounds. A synthesis method must discover feasible tests as well as select
enough of them to distinguish the candidates. Even under finite bounds, this
space may contain exponentially many different ways of dividing the
candidate set.

We address this challenge through two families of symbolic synthesis
methods. The \emph{cover-based} methods directly target minimum batch
cardinality. \jointcover{} jointly synthesizes tests and selects a minimum
cover of the candidate pairs in one maximum satisfiability (MaxSAT) query. A second
cover-based method, \enumeratedcover{}, uses repeated SAT queries to
enumerate the different ways feasible tests divide the candidates, retaining
one representative test for each
division, and then solves the resulting explicit minimum-cover problem with
MaxSAT. Both methods return a minimum-cardinality batch whenever they
complete, but differ in the
symbolic models they solve and the amount of information they materialize.

The \emph{tree-guided} methods adapt the recursive candidate-partitioning
principle of these interactive narrowing approaches to complete batch synthesis.
We develop two symbolic synthesis methods based
on this principle: \arbitrarysplit{} uses SAT to obtain any feasible split
and avoids optimization, whereas \balancedsplit{} uses MaxSAT to seek the
most balanced feasible split.
Both explore the two outcomes of every split before expert review and retain
generated tests for reuse across branches. Exploring both outcomes ensures
completeness; reuse allows a test synthesized for one candidate subset to
separate another subset without generating a new test. These methods provide
complete batches without guaranteeing minimum cardinality.

We formalize the batch-synthesis problem independently of a specification
language. We instantiate the four synthesis methods for Alloy's temporal relational
logic~\cite{jackson2012software,jackson2019alloy}. Our prototype uses
AlloyMax~\cite{zhang2021alloymax} for optimization and supports structural
instances and bounded temporal traces through an explicit lasso encoding. The
evaluation centers on two questions: how small the synthesized batches are,
including the contribution of reuse and the gap to the minimum; and how
synthesis time and completion depend on candidate count, structural scope,
and temporal bounds.

Our main benchmark comprises 196 static configurations derived from 28
requirements in three Alloy models, with candidate counts ranging from 4 to
28. A small temporal replication study adds 32 configurations from eight
requirements in one dynamic model with temporal properties, with 4 to 16 candidates. On 168 static
configurations completed by both tree-guided methods and at least one exact
method, test reuse reduces balanced batch cardinality by 42\% relative to the
unreused construction tree, while minimum batches are a further 29\% smaller.
\arbitrarysplit{} completes every static and temporal configuration in the main
study. The exact methods have different completion profiles: on the temporal
benchmark, \enumeratedcover{} completes 31 of 32 configurations, compared with 9
for \jointcover{}.

Complete batches also support adaptive review: a tool selects tests from the
batch as answers arrive and stops when one candidate remains. Our secondary
analysis shows that balanced batches nearly attain the theoretical minimum
number of classifications on average, while minimum-cardinality batches need
only slightly more. On completed static configurations, synthesizing the full
balanced batch costs a median of 0.16 seconds more than generating tests during
review. This additional upfront effort, however, makes a complete, reusable suite
available before any judgments are collected, supporting independent,
asynchronous, and adaptive review.

This paper contributes:
\begin{itemize}
  \item a formalization of complete distinguishing batches over an implicit
    symbolic test space, including their characterization as covers of
    candidate pairs;
  \item two exact synthesis methods that minimize batch cardinality through
    a single MaxSAT optimization or SAT enumeration followed by explicit
    minimum-cover solving with MaxSAT;
  \item two tree-guided synthesis methods, one using MaxSAT and the other using
    SAT, that adapt recursive candidate partitioning to complete batch
    construction with test reuse; and
  \item an evaluation of batch cardinality, synthesis cost, and scalability
    for structural instances and temporal traces, together with a secondary
    analysis of adaptive classification and the cost of complete
    precomputation.
\end{itemize}

\section{Illustrative Example}
\label{sec:example}

\begin{figure}[t]
\begin{alloy}[basicstyle=\scriptsize\ttfamily]
// domain model
sig Person { teaches, enrolled : set Course, workson : set Project }
sig Professor, Student in Person {}
sig Course { projects : set Project }
sig Project {}
fact Assumptions {
  all c : Course | some teaches.c
  teaches.Course in Professor and enrolled.Course in Student
  all p : Project | lone projects.p
  all p : Person | no (p.teaches & p.enrolled)
  all p : Person | p.workson in p.enrolled.projects
}
// candidate specifications
pred Spec1 { all p : Professor | no (p.teaches & (Professor - p).enrolled) }
pred Spec2 { all p,q : Professor |
 (some (p.enrolled & q.enrolled) and p!=q) implies no (p.teaches & q.enrolled)
}
pred Spec3 { all p,q : Professor |
 (some (p.teaches & q.teaches) and p!=q) implies no (p.teaches & q.enrolled)
}
pred Spec4 { all p : Professor |
 let colleagues = { q:Person | q!=p and some (p.workson & q.workson) } |
   no (p.teaches & colleagues.enrolled)
}
pred Spec5 { all p : Professor |
 let colleagues = { q:Professor | q!=p and some (p.workson & q.workson) } |
  no (p.teaches & colleagues.enrolled)
}
pred Spec6 { all p,q : Professor |
 (p!=q and some ((p.teaches.~enrolled) & (q.teaches.~enrolled))) implies
 no (p.teaches & q.enrolled)
}
\end{alloy}
\caption{University-course domain model and six non-equivalent candidate specifications returned by GPT-5.2 for a new requirement (analysis commands omitted).}
\label{fig:courses-model}
\end{figure}

Figure~\ref{fig:courses-model} shows an Alloy model of university courses.
Its signatures and fields represent people, courses, projects, teaching,
enrollment, and project work; the facts constrain their admissible
relationships. Suppose we want to formalize the new requirement ``a professor
cannot teach colleagues''. This requirement is ambiguous
because it does not define the concept of colleague. Asked to formalize it, GPT-5.2 returned
nine predicates; Fig.~\ref{fig:courses-model} shows the six that are
non-equivalent under the facts. They interpret colleagues as every other
professor (\a{Spec1}),
professors enrolled in a common course (\a{Spec2}), professors who co-teach a
course (\a{Spec3}), people who work on a common project (\a{Spec4}), professors
who work on a common project (\a{Spec5}), and professors who have a student in
common (\a{Spec6}). Alloy's relational notation and the similarity between
some predicates make direct comparison difficult. We
abbreviate \a{Spec}$i$ as $S_i$.

\paragraph{A complete distinguishing batch.}
Our goal is to synthesize a fixed suite that distinguishes all six candidates
before knowing which interpretation the expert intends.
Table~\ref{tab:batch-matrix} shows one possible three-test batch. A check mark
denotes satisfaction of a candidate specification by a test. The six columns
are distinct, so every candidate pair disagrees on at least one test. For
example, $S_2$ and $S_5$ both accept $T_1$ and $T_2$, but $T_3$ separates them.
Each row thus covers some of the $\binom{6}{2}=15$ candidate pairs, and the
three rows collectively cover all pairs. The batch is complete independently
of which candidate is intended.

\begin{table}[t]
\centering
\caption{A minimum-cardinality distinguishing batch for the six candidate
specifications. A check mark denotes satisfaction.}
\label{tab:batch-matrix}
\small
\setlength{\tabcolsep}{5pt}
\begin{tabular}{lcccccc}
\toprule
 & $S_1$ & $S_2$ & $S_3$ & $S_4$ & $S_5$ & $S_6$ \\
\midrule
$T_1$ & -- & \checkmark & \checkmark & -- & \checkmark & -- \\
$T_2$ & -- & \checkmark & -- & \checkmark & \checkmark & \checkmark \\
$T_3$ & -- & -- & \checkmark & \checkmark & \checkmark & -- \\
\bottomrule
\end{tabular}
\end{table}

\begin{figure}[t]
  \centering
  \includegraphics[width=.60\textwidth]{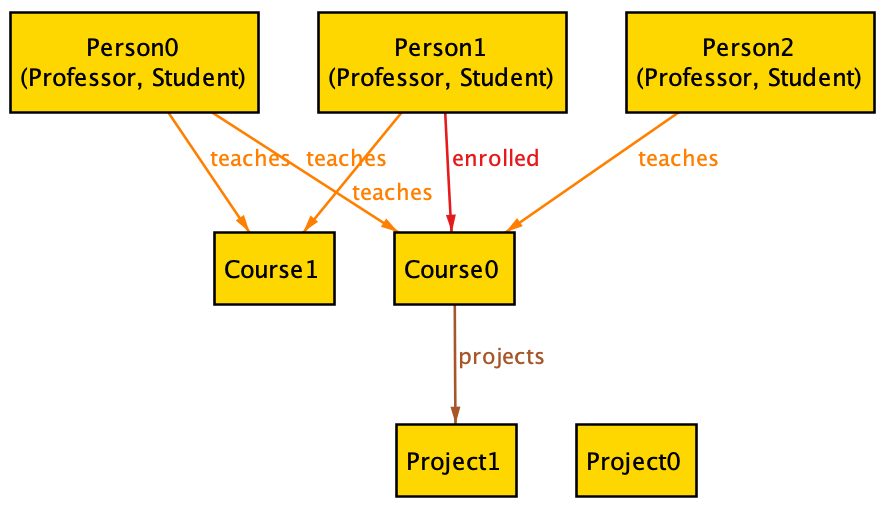}
  \caption{Test $T_2$ from the distinguishing batch, visualized as an Alloy instance.}
  \label{fig:courses-test2}
\end{figure}

\paragraph{Reviewing the complete batch.}
The three tests can be classified independently, in any order, or distributed
among reviewers. If the judgments for $T_1$, $T_2$, and $T_3$ are desirable,
undesirable, and desirable, respectively, they match only the column of $S_3$.
Reviewing the complete batch requires three classifications, regardless of
which candidate is ultimately selected.
Figure~\ref{fig:courses-test2} depicts $T_2$: one professor co-teaches a course
with another professor and teaches a course in which that colleague is
enrolled, precisely the situation rejected by \a{Spec3}. The expert can judge
this concrete scenario without reading the predicates.

\paragraph{Batch cardinality.}
Two binary tests encode at most four different acceptance patterns, so
distinguishing six candidates requires at least three tests. The batch in
Table~\ref{tab:batch-matrix} attains this bound and therefore has minimum
cardinality.
Table~\ref{tab:example-cardinality} compares the complete batches produced by
the four proposed synthesis methods on this example at scope~3. A complete binary
construction tree for six candidates has five internal test positions.
\arbitrarysplit{} stores five distinct tests in this run. \balancedsplit{} reuses
a test at one position and stores four distinct tests. \jointcover{} and
\enumeratedcover{} each produce a three-test minimum batch, although the tests in
their batches may differ. The example illustrates how reuse reduces the size
of a tree-guided batch, while global cover optimization can reduce it further.
Whether this reduction is worth the synthesis effort is a central question of
the evaluation.

\begin{table}[t]
\centering
\caption{Complete batch cardinalities for the six-candidate example at
scope~3. Both cover-based methods guarantee minimum cardinality.}
\label{tab:example-cardinality}
\small
\begin{tabular}{llr}
\toprule
Method & Family & Tests \\
\midrule
\jointcover{} & Cover-based & 3 \\
\enumeratedcover{} & Cover-based & 3 \\
\arbitrarysplit{} & Tree-guided & 5 \\
\balancedsplit{} & Tree-guided & 4 \\
\bottomrule
\end{tabular}
\end{table}

\paragraph{Interactive adaptive use.}
An interactive tool can also present the precomputed batch adaptively,
selecting an unused test that most evenly divides the remaining candidates.
Suppose $S_3$ is intended. In Table~\ref{tab:batch-matrix}, $T_1$ can be chosen
first; its desirable answer leaves $\{S_2,S_3,S_5\}$. Classifying $T_2$ as
undesirable then isolates $S_3$, so review can stop without classifying $T_3$.
Only two classifications are needed on this path, although the complete
batch contains three tests and was synthesized before either answer was
available. A secondary question in the evaluation is how the adaptive use of complete batches compares with online approaches.

\section{Distinguishing Test Batches}
\label{sec:problem}

\subsection{Candidates and Feasible Tests}

Let $F=\{\phi_1,\ldots,\phi_N\}$ be a finite set of candidate specifications
over a common vocabulary, and let $\mathcal I$ be the finite domain of tests
admitted by the background theory and analysis bounds. A test may be a static
instance or an execution trace; $I\models\phi$ means that test $I \in \mathcal I$ satisfies
candidate $\phi$. The domain $\mathcal I$ is represented implicitly by the
theory and bounds, rather than supplied as an explicit list of tests.
Two candidates are distinguishable within the bounds if some feasible test
satisfies exactly one of them. We assume throughout that the candidates are
pairwise distinguishable. Candidates that agree on all feasible tests cannot
be distinguished by test classification at the chosen bounds.

\subsection{Complete Batches and Minimum Covers}

\begin{definition}
A \emph{complete distinguishing batch} for $F$ is a finite set
$T\subseteq\mathcal I$ such that, for every two distinct candidates
$\phi_i,\phi_j\in F$, some $I\in T$ satisfies exactly one of them. A
distinguishing batch is \emph{minimum-cardinality} if no smaller such batch
exists.
\end{definition}

Equivalently, the candidates have distinct acceptance columns across $T$, as
illustrated in Table~\ref{tab:batch-matrix}. Completeness depends only on the batch and candidate set, independently of
the intended candidate and expert judgments.

Let $U=\{\{\phi_i,\phi_j\}\mid 1\leq i<j\leq N\}$ be the candidate pairs.
Each feasible test distinguishes
\[
  D(I)=\{\{\phi_i,\phi_j\}\in U\mid
          I\models\phi_i \Longleftrightarrow I\not\models\phi_j\}.
\]
A batch is complete exactly when its tests cover $U$. Minimum-cardinality
synthesis therefore solves
\[
  \min_{T\subseteq\mathcal I}|T|
  \qquad\text{subject to}\qquad
  \bigcup_{I\in T}D(I)=U.
\]
Given an explicit pool of tests, this is the classical NP-hard
\emph{Minimum Test Set} problem~\cite{GareyJohnson,moret1985minimizing}. Here, however,
the test domain is implicit: synthesis must discover feasible tests as well
as select a minimum cover. Even finite bounds may admit exponentially many candidate divisions.

\subsection{Cardinality Bounds and Synthesis Objectives}

\begin{proposition}
\label{prop:batch-bounds}
For a pairwise distinguishable set of $N>1$ candidates, every minimum-cardinality batch
$T^*$ satisfies
\[
  \lceil\log_2N\rceil\leq |T^*|\leq N-1.
\]
\end{proposition}

The lower bound follows because $k$ binary tests yield at most $2^k$
different acceptance columns. For the upper bound, start with all candidates
in one group and repeatedly choose a feasible test that splits a non-singleton
group, refining every group by its outcome. Pairwise distinguishability
ensures that such a test exists. Each step increases the number of nonempty
groups, so at most $N-1$ tests produce $N$ singleton groups and hence a
complete batch. Constraints on feasible tests may force the minimum above
the lower bound.

Our primary objectives are \emph{batch cardinality}, the number of distinct
stored tests $|T|$, and \emph{synthesis cost}, the total time required to
synthesize the complete batch. Full review requires exactly $|T|$ expert
classifications, so its classification count is already captured by
cardinality. Synthesis precedes expert review and is independent of the
intended candidate. Producing a smaller batch may require more synthesis
effort; we therefore compare methods by both batch cardinality and synthesis
time, including completion within a fixed budget as candidate counts and
analysis bounds increase.

\subsection{Reviewing a Synthesized Batch}

The tests in a complete batch can be classified independently and in any
order. A classification history is a sequence
$H=\langle I_1^{c_1},\ldots,I_k^{c_k}\rangle$, where $I_j\in T$ and
$c_j\in\{+,-\}$, with $+$ denoting desirable and $-$ undesirable. The
candidates consistent with that history are
\[
  C(H)=\{\phi\in F \mid
    I\models\phi \text{ for every } I^+\text{ in }H,\;
    I\not\models\phi \text{ for every } I^-\text{ in }H\}.
\]
After every test has been classified, completeness ensures $|C(H)|\leq 1$,
regardless of review order. A singleton identifies the candidate consistent
with all judgments; an empty set means that no candidate agrees with the
complete review.

\paragraph{Adaptive presentation.}
A precomputed batch can also be reviewed adaptively, stopping once at most one
candidate remains. For $C\subseteq F$, define
\[
  C_+(I)=\{\phi\in C\mid I\models\phi\},\qquad
  C_-(I)=\{\phi\in C\mid I\not\models\phi\}.
\]
A test \emph{splits} $C$ when both parts are nonempty. Let $C$ be the current
candidate set and $R\subseteq T$ the tests not yet classified. An adaptive
presentation policy $\rho$ maps every reachable state $(C,R)$ with $|C|>1$
to a test $\rho(C,R)\in R$ that splits $C$. For answer $c\in\{+,-\}$ to
$I=\rho(C,R)$, the next state is
\[
  \bigl(C_c(I),R\setminus\{I\}\bigr).
\]
A suitable test always exists: any two remaining candidates are distinguished
by a member of $T$, and previously classified tests cannot separate candidates
consistent with their answers. In Table~\ref{tab:batch-matrix}, for example,
$H=\langle T_1^+,T_2^-\rangle$ already gives $C(H)=\{S_3\}$, so adaptive
review can stop before $T_3$.

Our experiments use a greedy policy that minimizes
\[
  b_C(I)=\max\{|C_+(I)|,|C_-(I)|\},
\]
with deterministic tie-breaking. This policy aims for a short decision path
by ranking tests already in the batch. With a precomputed acceptance matrix,
selection and candidate updates require no further solving.

For the experimental analysis, we assume an intended candidate $\phi\in F$
serves as an \emph{oracle}, answering positively exactly when
$I\models\phi$. The \emph{adaptive classification cost}
$\mathcal C_\rho(F,T,\phi)$ is the number of requested tests before only the
oracle remains. Each splitting test eliminates at least one candidate, so
for $N>1$,
\[
  1\leq\mathcal C_\rho(F,T,\phi)\leq\min\{|T|,N-1\}.
\]
This cost is a secondary objective of our work. A minimum cover need not include the test
that most evenly splits a particular remaining set, while balanced splitting
may retain tests unnecessary for distinguishing the candidates as a whole.
The evaluation examines whether minimum-cardinality batches incur a material
adaptive classification penalty.

\section{Symbolic Test Synthesis}
\label{sec:synthesis}

The synthesis procedures require only a bounded propositional encoding of the
specification language. Let $X$ be the Boolean variables representing one test,
and let $B(X)$ characterize the feasible tests in $\mathcal I$, including the
background theory and analysis bounds. For each candidate $\phi_i$, let
$\widehat{\phi}_i(X)$ be true exactly when the test represented by $X$ satisfies
$\phi_i$. A satisfying assignment projected onto $X$ can therefore be decoded
into a concrete test. We write the encodings as arbitrary Boolean formulas;
conversion to CNF, including the introduction of auxiliary variables, is
standard.

Besides ordinary SAT, we use partial MaxSAT~\cite{cha1997local}, which maximizes
satisfied soft formulas subject to all hard formulas. This interface suffices
for every procedure below and is independent of the source language.
We specify each query by its set of hard constraints and, for MaxSAT, its
set of soft constraints.
We assume $N>1$; otherwise, the empty batch suffices.

\subsection{Cover-Based Batch Synthesis}
\label{sec:cover-based-synthesis}

The cover-based methods solve the minimum-cardinality problem from
Section~\ref{sec:problem}. Both require the selected tests collectively to
cover every candidate pair, differing in whether tests and their selection
are synthesized jointly or a finite pool of representatives is discovered
first.

\paragraph{\jointcover{}.}
Let $K=N-1$, the upper bound from Proposition~\ref{prop:batch-bounds}, and
create $K$ potential tests with disjoint Boolean variable copies $X^1,\ldots,X^K$.
Introduce validity variables $v_{k,i}$ for each test position and candidate,
and let $t_k$ indicate whether test position $k$ is selected. The hard constraints
form the set
\begin{equation}
\label{eq:joint-separation}
\begin{gathered}
  \{B(X^k)\mid 1\leq k\leq K\}
    \cup\{v_{k,i}\leftrightarrow\widehat{\phi}_i(X^k)
    \mid 1\leq k\leq K,\ 1\leq i\leq N\}\\
  \cup\\
  \Bigl\{\bigvee_{k=1}^{K}\bigl(t_k\land(v_{k,i}\oplus v_{k,j})\bigr)
    \;\Bigm|\;1\leq i<j\leq N\Bigr\}.
\end{gathered}
\end{equation}
The last constraint family requires some selected test to separate each candidate pair.
The soft constraints are
\begin{equation}
\label{eq:joint-objective}
  \{\neg t_k\mid 1\leq k\leq K\}.
\end{equation}
Maximizing their number minimizes the selected test positions in one MaxSAT
query. Equation~\eqref{eq:joint-separation} ensures completeness. Conversely,
Proposition~\ref{prop:batch-bounds} guarantees that every minimum batch fits
within the $K$ positions, so an optimum yields a minimum-cardinality batch.
Before auxiliary variables for CNF conversion, the encoding uses $K$ copies
of $X$, $KN$ validity variables, $K$ selection variables, and $O(KN^2)$
separation occurrences. Replicating the bounded test encoding can therefore
be expensive when individual tests are complex.

\paragraph{\enumeratedcover{}.}
This method separates test discovery from cover selection. Tests with the
same acceptance vector separate the same candidate pairs; complementary
vectors also have identical pair coverage. One representative per distinct
candidate division therefore suffices for minimum-cover selection.

Define the Boolean vector $p(X)\in\{0,1\}^{N-1}$ relative to $\phi_1$ by
\begin{equation}
\label{eq:separation-pattern}
  p(X)_i=1\quad\Longleftrightarrow\quad
    \bigl(\widehat{\phi}_{i+1}(X)
      \leftrightarrow\widehat{\phi}_1(X)\bigr),
  \qquad 1\leq i<N.
\end{equation}
For a concrete test $I$, write $p(I)$ for this vector evaluated on its
encoding. A bit is one when the corresponding candidate and $\phi_1$ agree
on the test. Thus, $p(I)$ records their relative agreement, independently of
which side accepts the test. The all-one vector $\mathbf 1$ represents
unanimous agreement and covers no pair.

After finding representatives $I_1,\ldots,I_m$, the next SAT query uses the
hard-constraint set
\begin{equation}
\label{eq:enumeration-query}
  \{B(X),\ p(X)\neq\mathbf 1\}
    \cup\{p(X)\neq p(I_r)\mid 1\leq r\leq m\}.
\end{equation}
Each satisfying assignment gives a representative of a previously unseen
nontrivial division. Its blocking condition excludes every test with the
same pair coverage, including complementary acceptance vectors. Unsatisfiability certifies that all feasible nontrivial divisions are
represented.

Let $M$ be the number of retained representatives, and introduce a selection
variable $y_r$ for each one. The explicit cover encoding has the
hard-constraint set
\begin{equation}
\label{eq:enumerated-cover}
  \Bigl\{\bigvee_{r:\{\phi_i,\phi_j\}\in D(I_r)}y_r
    \;\Bigm|\;\{\phi_i,\phi_j\}\in U\Bigr\},
\end{equation}
and the soft constraints
\begin{equation}
\label{eq:enumerated-objective}
  \{\neg y_r\mid 1\leq r\leq M\}.
\end{equation}
A MaxSAT optimum minimizes the selected representatives. Every test with
nonempty pair coverage can be replaced by its representative, and tests
covering no pair can be discarded.
These operations preserve completeness without increasing cardinality, so
the optimum is also a minimum over the implicit test domain.

The method uses $M+1$ SAT enumeration queries, including the final
unsatisfiable query, followed by one MaxSAT
cover query. Each enumeration query uses one copy of the bounded test
encoding, but $M$ can be as large as $2^{N-1}-1$. The number of queries and
the explicit cover can therefore grow exponentially with candidate count.

\subsection{Tree-Guided Batch Synthesis}
\label{sec:tree-guided-synthesis}

Both tree-guided methods start with a worklist containing $F$ and an empty
batch $T$. At each step, they remove a non-singleton candidate set $C$ and
select a feasible test that splits it, using a stored test or a new solver
query according to the method's reuse rule. A newly synthesized test selected
for the node is added to $T$, and both non-singleton parts are added to the
worklist. Stored tests can be reused only at nodes on other branches: a test
cannot split a descendant of a node it labels, since all candidates there
agree on its outcome. When the worklist is empty, $T$ contains the distinct
labels of the implicit binary construction tree.

\begin{proposition}
\label{prop:materialization}
For a complete tree-guided construction over $N>1$ pairwise distinguishable
candidates, the construction tree has exactly $N-1$ internal nodes and its
stored batch $T$ is distinguishing, with $|T|\leq N-1$.
\end{proposition}

Every internal node has two nonempty children and the leaves are the $N$
singleton candidate sets, giving $N-1$ internal nodes. Any two candidates
diverge at a node whose test distinguishes them, so the distinct node labels
cover every candidate pair. Reusing a label across nodes preserves this
coverage while reducing the number of stored tests.

\paragraph{\arbitrarysplit{}.}
At a node with candidate set $C$, this method reuses the first stored test
that splits $C$. If none exists, it asks a SAT solver to satisfy the
hard-constraint set
\begin{equation}
\label{eq:arbitrary-split}
  \Bigl\{B(X),\ \bigvee_{\phi_i\in C}\widehat{\phi}_i(X),\,
    \bigvee_{\phi_i\in C}\neg\widehat{\phi}_i(X)\Bigr\}.
\end{equation}
The last two constraints ensure that the split yields two nonempty sets. The method requires no optimization and gives no
minimum-cardinality guarantee.

\paragraph{\balancedsplit{}.}
This method maximizes the number of candidate pairs separated at the current
node. For each $\phi_i\in C$, introduce a validity variable $v_i$. The
hard-constraint set is
\begin{equation}
\label{eq:balanced-feasibility}
  \Bigl\{B(X),\ \bigvee_{\phi_i\in C}v_i,\,
    \bigvee_{\phi_i\in C}\neg v_i\Bigr\}
    \cup\{v_i\leftrightarrow\widehat{\phi}_i(X)\mid\phi_i\in C\}.
\end{equation}
The soft constraints are
\begin{equation}
\label{eq:balanced-objective}
  \{v_i\oplus v_j\mid \phi_i,\phi_j\in C,\ i<j\}.
\end{equation}
Each soft constraint corresponds to an unordered candidate pair and is
satisfied exactly when the test separates that pair. Every candidate in
$C_+(I)$ is separated from every candidate in $C_-(I)$, giving
$|C_+(I)|\,|C_-(I)|$ satisfied soft constraints. Since the two parts sum to
$|C|$, maximizing this product is equivalent to minimizing the larger part,
$b_C(I)$.

For $m=|C|$, the encoding adds $m$ validity variables and $\binom m2$ soft
constraints to one copy of the bounded test encoding. The optimization
concerns the current split; it guarantees neither minimum batch cardinality
nor globally minimum adaptive classification cost.

Reuse preserves this local optimum. We first evaluate the stored tests on
$C$. If one attains the ideal value $b_C(I)=\lceil |C|/2\rceil$, we reuse the
earliest such test without a synthesis query. Otherwise, a MaxSAT query
establishes the best feasible split. If a stored test matches that optimum,
we reuse the earliest such test and discard the new witness; otherwise the
new test is stored. Every added test is evaluated against all of $F$, making
its acceptance values available for reuse across branches. Ideal reuse saves a query and a stored test; non-ideal reuse saves only a
stored test.

\section{Evaluation}
\label{sec:evaluation}

Our evaluation addresses two main questions on complete batch synthesis and a
secondary comparison with adaptive iterative classification.

\textbf{RQ1 (Batch cardinality):} How small are the synthesized batches, how
much do reuse and minimum-cover optimization reduce their cardinality, and how
sensitive are these sizes to analysis scope?

\textbf{RQ2 (Synthesis cost and scalability):} What does synthesizing a complete
batch cost, and how do the methods scale with candidate count and analysis
bounds?

\textbf{RQ3 (Adaptive iterative classification):} How do complete batches
compare with iterative synthesis in classification effort and synthesis time?

\subsection{Implementation and Experimental Setup}
\label{sec:implementation}

\paragraph{Implementation.}
We implemented the four methods in a Python prototype using
AlloyMax~1.0.2~\cite{zhang2021alloymax}, with SAT4J for SAT and SAT4JMax for
partial MaxSAT. Given an Alloy domain model, candidate predicates, and analysis
bounds, the prototype returns tests as Alloy commands together with their
candidate truth vectors, obtained directly from the solver solutions.
Temporal formulas are encoded as bounded
lassos~\cite{biere1999symbolic}; the prototype currently excludes past operators
and mutable top-level signatures.

\paragraph{Benchmark.}
The main benchmark contains 28 requirements from three static Alloy4Fun
exercises: a photo-sharing social network, a production line, and a train
station~\cite{macedo2021experiences,alloy4fun_dataset,a4f_semantic_dataset}. For
each requirement, we select the instructor's oracle and representatives of the
$N-1$ most frequent incorrect semantic classes, using each class's most
frequent syntactic submission and excluding unsupported representatives.
Candidate counts $N\in\{4,8,12,16,20,24,28\}$ yield 196 configurations. A small
temporal replication study uses 8 requirements from Alloy4Fun's temporal
train-station model and $N\in\{4,8,12,16\}$, yielding 32 configurations. All
candidate sets are pairwise distinguishable at the default bounds. Sets
derived from the same requirement are nested, so our summaries are descriptive
rather than treating configurations as independent samples.

\paragraph{Protocol and measures.}

The default structural scope is~3, with maximum trace length~3 for temporal
models. Each complete batch has a 60-second synthesis budget. Experiments ran
serially on an Apple M5 machine with 24~GB of memory and macOS Tahoe, using
Python~3.14.6, JPype~1.7.1, and Java~17.0.5. We measure
stored-test cardinality and wall-clock time for complete batch synthesis.
Review simulation is excluded from synthesis time. RQ1 and RQ2 retain
successful syntheses; RQ3 requires completed review simulation.
Cardinalities are averaged across configurations. A minimum
cardinality is certified by either cover-based method that completes. Runtime
comparisons report medians on configurations completed by every method shown;
separately labelled completion diagnostics use each method's own successful runs.

For RQ3, every candidate in turn acts as the oracle, accepting exactly the
tests it satisfies. Batch review follows the greedy adaptive policy from
Section~\ref{sec:problem} and stops when only the oracle remains.
Classification means weight oracle paths equally. The iterative baseline uses
the same balanced splitting objective but synthesizes a new test after each
answer. We compare upfront synthesis of the complete batch with cumulative
iterative synthesis along each oracle path. We evaluate all iterative oracle
paths by traversing the decision tree once,
charging each oracle only for synthesis and partitioning along its own path.
Iterative path times are averaged per configuration before computing paired
runtime medians.

Sensitivity studies evaluate all four methods at structural scopes~3--7 for
static sets with $N=16$. Temporal
studies use $N=4$, varying either structural scope or maximum trace length from~3 to~7
while fixing the other bound at~3. RQ1 retains the same configurations with
both tree-guided batches and a minimum certified by either cover method at
every bound. RQ2 pairs configurations completed by all four methods at every
bound, explicitly identifying any reduced comparison subset. Full results
are available in the replication package.

\subsection{RQ1: Batch Cardinality}
\label{sec:rq1}

Table~\ref{tab:batch-cardinality} compares batch sizes on the 168 static
configurations completed by both tree-guided methods and at least one
cover-based method. Both cover-based methods certify the same minimum
cardinality whenever they complete, so their results share the minimum
column. All table means use this paired subset; RQ2 reports completion
separately.

\begin{table}[t]
\centering
\caption{Mean batch cardinality at scope~3 on paired static configurations.
$N-1$ is the number of construction-tree nodes before reuse.}
\label{tab:batch-cardinality}
\small
\setlength{\tabcolsep}{5pt}
\begin{tabular}{rrrrrr}
\toprule
$N$ & Configs & $N-1$ & \methodheading{Arbitrary}{Split} &
    \methodheading{Balanced}{Split} & Minimum \\
\midrule
4  & 28 &  3 &  2.82 &  2.71 & 2.46 \\
8  & 28 &  7 &  5.93 &  5.36 & 4.14 \\
12 & 28 & 11 &  7.86 &  7.18 & 5.14 \\
16 & 28 & 15 &  9.82 &  8.93 & 6.00 \\
20 & 25 & 19 & 11.24 & 10.28 & 6.96 \\
24 & 17 & 23 & 12.94 & 11.41 & 7.82 \\
28 & 14 & 27 & 14.21 & 12.07 & 8.71 \\
\midrule
All & 168 & 13.40 & 8.57 & 7.72 & 5.51 \\
\bottomrule
\end{tabular}
\end{table}

\paragraph{Reuse and minimum covers.}
The corresponding construction trees have 13.40 internal nodes on average.
Test reuse reduces the stored batches to 8.57 tests for \arbitrarysplit{} and
7.72 for \balancedsplit{}, reductions of 36\% and 42\%, respectively.
Minimum covers contain 5.51 tests on average, a further 29\% reduction
relative to \balancedsplit{}. On the 14 paired configurations with 28
candidates, fewer than nine tests on average suffice to distinguish every
candidate pair. Complete batch synthesis thus produces a compact suite for
full review even at the largest candidate count studied.

The small temporal replication shows the same reduction: on 30 paired
configurations, minimum batches average 4.27 tests compared with 6.00 for
\balancedsplit{}, again 29\% smaller.

\paragraph{Sensitivity to analysis bounds.}
For $N=16$, the same 23 configurations have both tree-guided batches and a
certified minimum at every scope from~3 to~7. Across these scopes,
\balancedsplit{} averages 8.74--9.13 tests and \arbitrarysplit{} averages
9.78--10.35. The mean minimum, certified by either cover method,
decreases from 5.96 tests at scope~3 to 5.83 at
scope~4 and remains unchanged through scope~7. These comparisons hold the
candidate sets and completed subset fixed.
In the paired static scope study, scope~3 already achieves nearly the same
mean minimum batch cardinality as scopes~4--7. This finding is consistent
with Alloy's \emph{small scope hypothesis}~\cite{jackson2012software} and
supports the Analyzer's default scope of~3 as a practical starting point for
complete batch synthesis.

On seven paired temporal configurations with $N=4$ and maximum trace length~3,
the mean minimum remains 2.29 tests across structural scopes~3--7; tree-guided means range
from 2.43 to 3.00. When varying the trace-length from 3 to 7 with fixed structural scope~3, the mean minimum remains
2.25 tests on eight paired configurations,
with tree-guided means of 2.62--3.00. RQ2 examines the synthesis cost of
increasing these bounds. Increasing either analysis bound leaves the mean minimum batch cardinality unchanged on the paired temporal subsets.

\subsection{RQ2: Synthesis Cost and Scalability}
\label{sec:rq2}

Table~\ref{tab:synthesis-cost} reports completion at the default bounds and
median synthesis times on the 133 static configurations completed by all four
methods. Completion counts cover the full benchmark; the paired medians
describe successful runs on a common subset.

\begin{table}[t]
\centering
\caption{Completion at default bounds under the 60-second budget. The final
row gives median synthesis times on the same 133 static configurations
completed by all four methods.}
\label{tab:synthesis-cost}
\small
\setlength{\tabcolsep}{5pt}
\begin{tabular}{lrrrrr}
\toprule
$N$/benchmark & Configs & \methodheading{Arbitrary}{Split} &
    \methodheading{Balanced}{Split} & \methodheading{Joint}{Cover} &
    \methodheading{Enumerated}{Cover} \\
\midrule
4  & 28 & 28 & 28 & 28 & 28 \\
8  & 28 & 28 & 28 & 28 & 28 \\
12 & 28 & 28 & 28 & 26 & 28 \\
16 & 28 & 28 & 28 & 24 & 24 \\
20 & 28 & 28 & 27 & 24 & 20 \\
24 & 28 & 28 & 21 & 19 & 14 \\
28 & 28 & 28 & 17 & 15 & 11 \\
\midrule
Static total & 196 & 196 & 177 & 164 & 153 \\
Temporal total & 32 & 32 & 31 & 9 & 31 \\
\midrule
\multicolumn{2}{l}{Static median (s)} & 0.18 & 0.42 & 1.06 & 0.57 \\
\bottomrule
\end{tabular}
\end{table}

\paragraph{Completion and candidate count.}
\arbitrarysplit{} completed all 228 static and temporal configurations at
the default bounds, each in at most 0.43~s. \balancedsplit{} completed all
static configurations through $N=16$, but only 17 of 28 at $N=28$. The
cover-based methods also become less robust as candidate count increases:
at $N=28$, \jointcover{} completed 15 configurations and
\enumeratedcover{} completed 11. Avoiding optimization therefore provides
the most reliable complete batches in this benchmark, at the cost of the
larger batch cardinalities reported in RQ1.

\paragraph{Cost of minimum covers.}
On the common static subset, all four median synthesis times are below
1.1~s. \arbitrarysplit{} is fastest at 0.18~s, followed by
\balancedsplit{} at 0.42~s. Minimum covers remain inexpensive on this subset:
\enumeratedcover{} takes 0.57~s, compared with 1.06~s for \jointcover{}.
However, these medians do not capture the harder configurations that time
out. \enumeratedcover{} discovers a median of 33 distinct candidate
divisions (maximum 278) on its completed static runs; the reported time
includes both SAT enumeration and MaxSAT cover selection. Its exhaustive
discovery phase limits scalability as candidate count grows.

\paragraph{Temporal replication.}
\enumeratedcover{} completed 31 of 32 temporal configurations, compared
with nine for \jointcover{}. This difference is consistent with the cost
of replicating the bounded-lasso encoding up to $N-1$ times in
\jointcover{}, whereas each enumeration query encodes only one test.
The temporal sets also contain at most 16 candidates, so this small study
does not isolate the effect of temporal encoding from candidate count.
Enumeration therefore appears attractive for expensive test encodings and
moderate candidate sets.

\paragraph{Sensitivity to analysis bounds.}
On the 11 static configurations with $N=16$ completed by all four methods
at every scope, increasing scope from~3 to~7 raises median synthesis time
from 0.18 to 0.28~s for \arbitrarysplit{}, from 0.58 to 1.49~s for
\balancedsplit{}, from 1.50 to 18.79~s for \jointcover{}, and from
1.47 to 3.33~s for \enumeratedcover{}. Across all 28 configurations,
\jointcover{} completion falls from 24 to 14 and \enumeratedcover{} from
24 to 21; at scope~7, \arbitrarysplit{} completes all 28 configurations
and \balancedsplit{} completes 26.

Increasing temporal trace length from~3 to~7 also raises synthesis time:
on the seven configurations completed by all four methods at every bound
with $N=4$, \balancedsplit{} rises from 0.89 to 2.67~s and
\jointcover{} from 1.05 to 3.43~s, while \arbitrarysplit{} and
\enumeratedcover{} remain below 0.4~s.
For temporal structural scope, six configurations complete all four methods
at every bound. On this common subset, increasing scope from~3 to~7 raises
median synthesis time from 0.13 to 0.22~s for \arbitrarysplit{}, from
0.86 to 8.25~s for \balancedsplit{}, from 0.97 to 13.46~s for
\jointcover{}, and from 0.20 to 0.38~s for \enumeratedcover{}.
At scope~7, \arbitrarysplit{} and \enumeratedcover{} complete all eight
configurations, compared with seven for \balancedsplit{} and six for
\jointcover{}. Larger bounds can therefore substantially increase synthesis
effort, particularly for MaxSAT over symbolic test encodings, with little
reduction in complete batch cardinality.

\subsection{RQ3: Adaptive Iterative Classification}
\label{sec:rq3}

\paragraph{Classification effort.}
On the 177 static configurations completed by \balancedsplit{} (2,640
oracle paths), adaptive batch review and the iterative baseline both average
4.16 classifications, differing by less than 0.001. For $N$ equally weighted
candidates, $B(N)=k+2-2^{k+1}/N$, with $k=\lfloor\log_2N\rfloor$, is the
minimum mean depth of a binary decision tree, even if arbitrary binary
questions are available. The batch mean is only 0.0038 above this
path-weighted lower bound. On this subset, even an optimal iterative policy
could therefore save fewer than 0.004 classifications on average.

On the 133 static configurations completed by all four methods (1,652
oracle paths), the other three methods require only 0.031--0.035 more
classifications than the iterative baseline's mean of 3.896, despite their
different batch cardinalities. In the small temporal replication, the
\balancedsplit{} batch and iterative baseline have the same mean of 3.48
classifications across 304 oracle paths from 31 configurations. Complete
batches, including minimum covers, thus remain competitive with adaptive
iterative classification on these completed benchmarks.

\paragraph{Upfront synthesis cost.}
Median synthesis time is 0.61~s for the complete \balancedsplit{} batch, compared with
0.43~s for cumulative iterative synthesis along an oracle path on the 177
static configurations. Iterative times average oracle-path costs within each
configuration; the median paired increase for complete batch synthesis is
0.16~s. On the 31 temporal
configurations, the corresponding medians are 3.11 and 2.29~s,
with a median paired increase of 0.82~s.
These comparisons exclude the batch timeouts reported in RQ2.
Iterative synthesis is already fast on these benchmarks, so these results do not
establish a substantial reduction in review delays. The additional upfront effort,
however, provides a complete reusable suite before review, with adaptive presentation
requiring no further synthesis.

\subsection{Threats to Validity}
The educational Alloy models, nested candidate sets, and selection of frequent,
supported semantic classes from student submissions limit generalization to
industrial or GenAI-generated candidate pools. The temporal replication
covers only one model. Minimum cardinalities and observed scope stability
are relative to the analyzed bounds and fixed candidate sets. Cardinality
measures the number of tests; expert comprehension and review time remain
unmeasured. RQ3 uses one balanced iterative baseline under the same encoding,
weights candidates equally as oracles, and assumes that the intended
candidate is present and answers are error-free. Synthesis times depend on
the encoding, solver versions, and hardware. Paired cardinality and runtime summaries condition
on completion and can favor easier configurations. We use one toolchain,
retain common comparison subsets, and report completion separately.

\section{Related Work}
\label{sec:related}

\paragraph{Minimum test sets and suite synthesis.}
Given an explicit finite test pool, Minimum Test Set selects a smallest subset
that distinguishes objects by their test
outcomes~\cite{GareyJohnson,moret1985minimizing}. Our cover-based methods extend
this objective to an implicit symbolic test space, where feasible tests must
also be discovered.

Fault-based test generation also synthesizes distinguishing tests. Gargantini
and Fraser~\cite{gargantini2011generating} generate compact suites for Boolean
specifications using SAT/SMT, with test reuse and heuristic suite reduction.
Arcaini et al.~\cite{arcaini2015generating} apply these techniques to feature
models and propose user review of distinguishing configurations for model
validation. Their adequacy criterion distinguishes a reference specification
from its mutants. Our criterion separates every candidate pair, and our
cover-based methods guarantee minimum cardinality over the bounded test domain.

Related suite optimization targets program coverage. Test-suite reduction
retains existing tests that preserve a chosen coverage
criterion~\cite{khan2018systematic,hsu2009mints}. Whole-suite generation
optimizes coverage and suite size~\cite{fraser2012whole}. SAT-based
approaches generate minimum suites for modified condition\slash decision
coverage~\cite{kitamura2018optimal,yang2018generating}. Our coverage elements
are candidate pairs, and feasible tests are defined by the background theory
and analysis bounds.

\paragraph{Interactive narrowing.}
Shriver et al.~\cite{shriver2017end} narrow program candidates using generated
inputs whose expected outputs are supplied by the user.
LTLTalk~\cite{gavran2020interactive} similarly generates distinguishing
behaviors to refine temporal specifications. ARTEMIS, by Mendoza et
al.~\cite{mendoza2026automating}, seeks balanced distinguishing traces to
reduce the number of user judgments. PICK~\cite{prasad2026pick} presents
distinguishing scenarios for GenAI-generated regular expressions,
access-control policies, and temporal formulas, incorporating uncertain
answers and elimination thresholds and evaluating the workflow through user
studies.

These approaches use feedback to guide subsequent question generation. Our
tree-guided methods adapt the recursive partitioning principle to complete
batch synthesis: they explore both outcomes before review and reuse stored
tests across branches. The resulting objective concerns the number of distinct
stored tests and the synthesis effort needed to prepare the entire suite.

\paragraph{Decision trees and state identification.}
The expected number of questions is a standard objective for decision trees;
minimizing it is NP-hard~\cite{hyafil1976constructing}. Generalized Binary Search greedily
chooses a query that most evenly splits the remaining
hypotheses~\cite{nowak2011geometry}. This criterion underlies the local
objective of \balancedsplit{}, but local balancing does not minimize the
number of distinct stored tests.

FSM state identification uses observed outputs to identify an unknown state
through preset or adaptive distinguishing sequences. Hierons and
T\"urker~\cite{hierons2015incomplete} study small collections of incomplete
sequences that collectively distinguish all states. This relates to collective
coverage by a batch; in our setting, the outcomes are expert judgments on
instances or traces of competing specifications.

\paragraph{Tests from formal specifications.}
AUnit~\cite{sullivan2017automated} generates Alloy tests for coverage and
mutation analysis. TestEra~\cite{marinov2001testera} uses Alloy specifications
to generate tests for Java programs, while model-based
testing~\cite{dalal1999model} derives tests from system models. Our adequacy
criterion requires a collection of tests that distinguishes every pair in a
given candidate set.

\paragraph{LLM-generated validation tests.}
Cunha and Macedo~\cite{cunha2025validatingformalspecificationsllmgenerated}
evaluate the generation of positive and negative Alloy tests from natural
language requirements. Such tests can filter candidate specifications by
checking their agreement with the expected outcomes, but several distinct
candidates may still pass the same suite. The techniques are complementary:
after initial filtering with LLM-generated tests, our methods could synthesize
a complete distinguishing batch for the surviving candidates. Expert review
of this additional batch would then resolve the distinctions left open by the
initial suite.

\section{Conclusion}
\label{sec:conclusion}

We formalized complete distinguishing batches as covers of candidate pairs
and developed four symbolic synthesis methods, instantiated for Alloy
instances and bounded temporal traces. The cover-based methods minimize
cardinality, while the tree-guided methods reduce stored tests through reuse
across branches. Complete batches provide a fixed suite before expert review,
make the full review workload explicit upfront, and support independent or
parallel classification.

On 168 paired static configurations, reuse reduces \balancedsplit{} batch
cardinality by 42\%, and minimum covers are a further 29\% smaller. With 28
candidates, minimum covers average 8.71 tests on the 14 paired configurations.
In the paired static scope study, scope~3 already yields nearly the same
minimum cardinality as larger scopes, which require substantially more
synthesis effort. \arbitrarysplit{} completes all 196 static and 32 temporal
configurations at the default bounds, while the cover-based methods have
different completion profiles, particularly for temporal models.

Adaptive presentation of pre-computed batches remains competitive with iterative narrowing: balanced
batches nearly attain the theoretical lower bound on mean classification
cost, and minimum covers incur only a small additional cost. The extra
upfront synthesis effort produces a complete reusable suite whose adaptive
review requires no further synthesis.

Future work will extend the evaluation to realistic case studies, particularly
those with larger sets of temporal requirements. User studies will assess
expert review effort and guide the tuning of optimization criteria to account
for the difficulty of understanding and classifying individual tests.

\subsubsection*{Acknowledgements.}
OpenAI's GPT-5 and GPT-6 models were used to assist with coding, analysis of
experimental results, and manuscript preparation. The research ideas are the
authors' own. The authors thoroughly reviewed the resulting code, analyses,
and text and take full responsibility for the final work.

\section*{Data Availability}
The implementation, benchmark models, complete evaluation results, and reproduction instructions is available
at \url{https://github.com/haslab/Specification-Narrowing}.

\bibliographystyle{splncs04}
\bibliography{bibliography}

\end{document}